\documentclass[11pt,a4paper,english,nofootinbib]{revtex4}
\usepackage{lmodern}

\usepackage[T1]{fontenc}
\usepackage[utf8]{inputenc}
\usepackage{babel}
\usepackage{amsmath}
\usepackage{graphicx}
\usepackage{amssymb}
\usepackage[unicode=true, pdfusetitle,
 bookmarks=true,bookmarksnumbered=false,bookmarksopen=false,
 breaklinks=false,pdfborder={0 0 1},backref=false,colorlinks=false]
 {hyperref}
\usepackage{amsmath}
\usepackage{amssymb}
\def\b{\begin{equation}}
 \def\e{\end{equation}}

\newcommand{\dif}{\mathrm{d}}
\newcommand{\bq}{\begin{eqnarray*}}
\newcommand{\eq}{\end{eqnarray*}}
\newcommand{\beq}{\begin{eqnarray}}
\newcommand{\enq}{\end{eqnarray}}

\makeatletter
\usepackage{latexsym}\usepackage{bm}

\begin{document}

\title{ Particle Creation in some LRS Bianchi I models}
\author{Luis O. Pimentel  }
\email{lopr@xanum.uam.mx (Corresponding author)}
\author{Flavio Pineda}
\email{fpineda@xanum.uam.mx }
\affiliation{Physics Department, Universidad Aut\'onoma Metropolitana Iztapalapa P. O. Box 55-534, 09340 M\'exico, CDMX., M\'exico}
\date{\today}

\begin{abstract}

\noindent \hspace{0.35cm} 
In this work we consider particle creation by the expansion  of the universe, using two Bianchi type I anisotropic models.
The particles studied are of spin 0 and 1/2. The cosmological models have rotational symmetry, which allows us to solve 
exactly the equations of motion. The number density of the created  particles  is calculated with the method of  Bogolubov transformations.
\end{abstract}

\maketitle

\section{Introduction }

Quantum effects of the gravitational field are one of the greatest mysteries of nature; having a quantum theory of gravity would be one of the greatest achievements of modern science. However, we do not currently have a satisfactory theory of the quantum nature of the gravitational field that can describe and explaining the various scenarios of the very early universe. 
Although we do not have a satisfactory quantum gravity theory, there is no impediment to develop quantum field theory in non-flat space-time; quantum field theory can be developed including gravitational fields without a quantum theory of gravity,using a  classical  gravitational field, i.e. one given as a  Lorentzian metric that is a solution of Einstein’s equations.

The scheme is a theory describing the dynamics of quantum fields propagating in a curved space-time background, described by a Lorentzian manifold with a general classical metric $g_{\mu\nu}$. In this way it is possible to go quite far in generalizing quantum field theory without considering the difficulties and problems involved in any quantum field theory of gravity. The Unruh effect, Hawking radiation, the production of particles in the early universe, the generation of primordial gravitational waves or even the explanation of the isotropy of the universe are some consequences of the quantum field theory in curved space-time. Particle production from vacuum is one of the most amazing predictions of quantum theory; in curved space-time, production takes place because of very intense or changing gravitational fields such as the expansion of the universe, an effect Schrödinger expected to occur \citep{sch6}, or the field produced by a black hole, effect studied by Hawking \citep{haw4}.  Parker’s pioneering works \citep{par7, par8, par9} establishes that there is a creation of particles in the very early stages of the expansion of the universe; if this particle creation at very early times is characterized by the fundamental constants  $\hbar\,, c\,, G$, then is consistent with the entropy demanded by the cosmic background radiation with a temperature of $2,7 \, \mathrm{K}$ \citep{par10, par11}.

The majority of the  works on particle production deal with the production of particles in homogeneous and isotropic universes without initial singularity in $t=0$, for example, De Sitter’s universe \citep{Villa1}, models of  Friedman-Robertson-Walker \citep{Grib, Audretsch} or models of an asymptotically flat universe \citep{Moradi}; very few works discuss the production of particles in homogeneous anisotropic universes with initial singularity. The reason for this is partly that the usual techniques of quantum field theory in gravitational backgrounds fail when there is an initial singularity, so a different approach must be taken. 

The different methods that exist to address this problem are

\begin{enumerate}
\item Hamiltonian diagonalization \citep{Grib, Buk}
\item Feynman path-integral method \citep{Duru, Hartle}
\item  Semi-classical method \citep{Villa1, Villa2, Villa3}
\end{enumerate}

Bukhbinder \citep{Buk} develops the Hamiltonian diagonalization method and applies it to calculate the mean number of scalar particles created in a Bianchi I anisotropic universe, and the result obtained is  Bose’s distribution; Chitre and Hartle \citep{Hartle} developed the path integral method as a quantization method for a scalar field propagating in a homogeneous universe with a linear expansion, $a(t)=t$ and initial singularity in $t=0$, while Duru \citep{Duru} uses this method to calculate the mean number of scalar particles in  Bianchi I model	

\begin{equation}
\dif s^2=-\dif t^2+t^{2q}\,(\dif x^2+\dif y^2)+t^{2(1-2q)}\,\dif z^2\,.
\label{eq1}
\end{equation}

This metric is a one-parameter family of solutions to Einstein's equations with a   perfect stiff fluid or a free massless scalar field as material content\citep{Jacobs}.	

The semi-classical method has been used in different scenarios with favourable results, in particular, Villalba \citep{Villa1, Villa2, Villa3} has used it to discuss particle production in different scenarios. This method consists of solving the covariant Hamilton-Jacobi equation and compares the asymptotic behaviour of the solutions with the asymptotic behaviour of the solutions of the main field equations (Klein-Gordon equation or Dirac equation). In this paper, we discuss the production of scalar and spin 1/2 particles in some Bianchi I LRS models by the semi-classical method. We present the asymptotic solutions of the field equations and compared them with the exact solutions of the Klein-Gordon (KG) and Dirac equations for massive and non-massive particles, to define the positive and negative frequency states; to solve the Dirac equation, we used the method of separation of variables developed by Villalba and Shihskin \citep{shishkin1}. Finally, we calculated the number density of the created particles  with the method of the Bogolubov transformations.

 The cases that interest us are the following:

\beq
q=0\,,\quad \dif s^2 &=&-\dif t^2+\dif x^2+\dif y^2+t^2\,\dif z^2\\[0,2cm]
q=1\,,\quad\dif s^2 &=&-\dif t^2+t^2\,(\dif x^2+\dif y^2)+t^{-2}\,\dif z^2\,.
\enq

The case $q=0$ represents flat space-time, which is a particular solution of the Kasner metric, while the case $q=1$ represents an expanding homogeneous universe.

\section{Asymptotic Solution to the Hamilton Jacobi Equation}

We must solve the Hamilton-Jacobi equation for the  LRS Bianchi I model given by the metric \eqref{eq1}. To define particles with the exact solution of the KG equation a generally covariant approach should be used, which is based on solutions of the Hamilton-Jacobi equation in the respective space-time. The way to proceed consists of the following steps \citep{Audretsch}

\begin{enumerate}
\item  Solve the Hamilton-Jacobi equation for the respective cosmological model.

\item Solve the KG and Dirac equations for the respective cosmological model.

\item 	Compare the asymptotic behaviour of the solutions of the field equations and the Hamilton-Jacobi equation.

\item Positive and negative frequency states are obtained 
 according to the following correspondence 
 
\beq
\Phi(x)\sim\left\{ \begin{array}{rcl}
e^{-iS} & \mbox{Positive frequency states}\label{correspondecia WKB}\\[0.3cm]
e^{iS} & \mbox{Negative frequency states}\,,
\end{array}
\right.
\enq 
  
 where $S$ is the classical action and $\Phi$ is the wave function.
 
\end{enumerate}

The  covariant Hamilton-Jacobi equation can be written as

\beq
g^{\mu\nu}\,\partial_\mu S\,\partial_\nu S+m^2=0\,.\
\label{HJ}
\enq

Since the metric \eqref{eq1} is only dependent on time, the variables of the $S$ function can be separated as

\beq
S(\mathbf{k}\,,\mathbf{r}\,,t)=\mathbf{k}\cdot\mathbf{r}+f(t)\,.
\label{eq 2}
\enq

Substituting \eqref{eq 2} into \eqref{HJ} we obtain the equation for the $f$ function

\beq
-\left(\dfrac{\dif f}{\dif t}\right)^2+t^{-2q}\,k_\perp^2 +t^{2\,(2q-1)}\,k_z^2+m^2=0\,,
\enq

where $k_\perp^2=k_x^2+k_y^2$. For $q=0$, the asymptotic behavior of the solution is

\beq
f_k(t)\sim\pm \,\log t^{k_z}\,,
\enq

as $t\to 0$, that is, in the initial singularity, and

\beq
f_k(t)\sim \pm \sqrt{k_\perp^2+m^2}\,t\,,
\enq

as $t\to \infty$.  The classic action for this Bianchi model has the following asymptotic behaviour

\beq
S(\mathbf{k}\,,\mathbf{r}\,,t)\sim\left\{\begin{array}{rcl}
&\mathbf{k}\cdot\mathbf{r} \pm\log t^{k_z}\,,\quad t\to 0\label{acción 0 asintótico}\\[0.3cm]
&\mathbf{k}\cdot\mathbf{r} \pm\sqrt{k_\perp^2+m^2}\,t\,,\quad t\to \infty\,.
\end{array}
\right.
\enq

Therefore, the wave function $\Phi_k$ should have the following semiclassical behaviour

\beq
\Phi_k\sim\left\{\begin{array}{rcl}
e^{i\,\mathbf{k}\cdot\mathbf{r}}\,t^{\pm i\,k_z}\,,\quad t\to 0\label{campo WKB 0}\\[0.3cm]
e^{i\,\mathbf{k}\cdot\mathbf{r}}\,e^{\pm i\sqrt{k_\perp^2+m^2}\,t}\,,\quad t\to\infty\,.
\end{array}
\right.
\enq

For the case $q=1$, the asymptotic behaviour of the solution of the Hamilton-Jacobi equation is

\beq
f_k(t)\sim \pm\log t^{k_\perp}\,,
\enq

as $t\to 0$ and 

\beq
f_k(t)\sim \pm\dfrac{k_z\,t^2}{2}\pm\dfrac{k_z\,m^2}{2}\log t\,,
\enq

as $t\to \infty$. Hence,  the wave function $\Phi$ should have the following semi-classical behaviour

\beq
\Phi_k\sim\left\{\begin{array}{rcl}
e^{i\,\mathbf{k}\cdot\mathbf{r}}\,t^{\pm i\,k_\perp}\,,\quad t\to 0\label{campo WKB}\\[0.3cm]
e^{i\,\mathbf{k}\cdot\mathbf{r}}\,e^{\pm i\,k_z\,t^2/2}\,t^{\pm i\,k_z\,m^2/2}\,,\quad t\to\infty\,.
\end{array}
\right.
\enq

To identify the negative and positive frequency states of scalar and spin 1/2 particles it is necessary to solve the KG and Dirac equation in the Bianchi I model for cases $q=0$ and $q=1$.

\section{Klein Gordon Equation}

The massive KG equation with arbitrary coupling $\xi$ in curved space-time takes the form

\beq
(g^{\mu\nu}\,\nabla_\mu\,\nabla_\nu-m^2-\xi\,R)\Phi=0\,,
\enq

where $R$ is the scalar curvature, $\xi$ is a dimensionless coupling constant, which in the case of a conformally coupled field takes the value of $\xi=1/6$ and for the minimal coupling case $\xi=0$, and $\nabla_\mu$ is the covariant derivative. For the metric \eqref{eq1}, the equation is

 \beq
\dfrac{\dif^2 f_k}{\dif t^2}+\left[\dfrac{1}{t^2}\left(\dfrac{1}{4}+2q\,\xi(3q-2)\right)+t^{-2q}\,k_\perp{}^2+t^{4q-2}\,k_z{}^2+m^2\right]f_k(t)=0\,,
\label{KG simplificada}
\enq

where we have separated variables of the form 

\beq
\Phi(x)=t^{-1/2}\,e^{i\,\mathbf{k}\cdot\mathbf{r}}\,f_k(t)\,,
\enq

and we have substituted the scalar curvature for the metric \eqref{eq1}

\beq
R=\dfrac{2q\,(3q-2)}{t^2}\,.
\enq

Exact solutions for the cases $q=0$ and $q=1$ were obtained by Pimentel \citep{pim1}.

\begin{itemize}

\item  For the case $q=0$, the KG equation takes the form

\beq
\dfrac{\dif^2 f_k}{\dif t^2}+\left[\dfrac{1}{t^2}\left(\dfrac{1}{4}+k_z{}^2\right)+k_\perp{}^2+m^2\right]f_k(t)=0\,,
\enq

and the solution is

\beq
\Phi(x)=e^{i\,\mathrm{k}\cdot\mathrm{r}}\,[A_1\,H_{i\,k_z}^{(1)}(\sqrt{k_\perp{}^2+m^2}\,t)+B_1\,H_{i\,k_z}^{(2)}(\sqrt{k_\perp{}^2+m^2}\,t)]\,,
\label{campo escalar 0}
\enq

where $H_{\nu}^{(j)}$ is the Hankel function of order $\nu$ and $A_1\,, B_1$ are integration constants.

\item For the case $q=1$, the KG equation takes the following form:

\beq
\dfrac{\dif^2 f_k}{\dif t^2}+\left[\dfrac{1}{t^2}\,\left(\dfrac{1}{4}+2\xi+k_\perp{}^2\right)+t^2\,k_z{}^2+m^2\right]f_k(t)=0\,,
\enq

and the solution is

\beq
  \Phi(x)=t^{-1}\,e^{i\,\mathbf{k}\cdot\mathbf{r}}\,[A_1\,M_{\kappa\,,\mu}(ik_z\,t^2)+B_1\,W_{\kappa\,,\mu}(ik_z\,t^2)]\,,
  \label{campo escalar q=1}
   \enq 
 
 where $M_{\kappa\,\mu}$, $W_{\kappa\nu}$ are Whittaker's function and
 
  \beq
 \kappa=-\dfrac{im^2}{4\,k_z}\,,\quad \mu= \dfrac{i}{2}\,\sqrt{k_\perp{}^2+2\xi}\,.
 \enq

\end{itemize}

\section{Production of  Scalar Particles }

To identify the positive and negative frequency states in the $t\to 0$ and $t\to\infty$, the asymptotic solutions should be compared with the solution obtained from the semiclassic method.

\begin{enumerate}
\item Case $q=0$.

The solution of KG's equation for $q=0$ are Hankel functions \eqref{campo escalar 0}; the asymptotic behaviour of the Bessel function $J_\nu(z)$ for  $z\to 0$ is given by the asymptotic formula \citep{NIST}

\beq
J_\nu(z)\sim \dfrac{z^\nu}{2^\nu\,\Gamma(1+\nu)}\,.
\label{asintotico bessel 0}
\enq

Therefore, the solution \eqref{campo escalar 0} presents the following asymptotic behaviour in $t\to 0$

\beq
J_{i\,k_z}(\sqrt{k_\perp{}^2+m^2}\,t)\sim \dfrac{t^{i\,k_z}}{2^{i\,k_z}\,\Gamma(1+i\,k_z)}\sim t^{i\,k_z}\,,  \quad t\to0\,.
\enq

	Compared to the solution of the semiclassic method \eqref{campo WKB 0}, we see that the states of negative frequency are defined as
	
	\beq
f_{(0)}^-(t)=A_{(0)}^-\,J_{i\,k_z}(\sqrt{k_\perp{}^2+m^2}\,t)\,,
\label{frecuencia negativa q=0 cero}
\enq
 where $A_{(0)}^-$ is a normalization constant. The positive frequency state $f_{(0)}^+(t)$ is the complex conjugate of \eqref{frecuencia negativa q=0 cero}

\beq
f_{(0)}^+(t)=[f_{(0)}^-(t)]^*=A_{(0)}^+\,J_{-i\,k_z}\,(\sqrt{k_\perp{}^2+m^2}\,t)\,.
\label{frecuencia positiva q=0 cero}
\enq

On the other hand, the asymptotic behaviour of the Hankel function $H_{\nu}^{(2)}(z)$ for $z\to \infty$ is \citep{NIST}

\beq
H_\nu^{(2)}(z)\sim \sqrt{\dfrac{2}{\pi\,z}}\exp[-i(z-\nu\,\pi/2-\pi/4)]\,.
\label{asintótico Hankel}
\enq

Therefore, the solution \eqref{campo escalar 0} presents the following asymptotic behaviour in $t\to \infty$
 
\beq
H_{i\,k_z}^{(2)}(\sqrt{k_\perp{}^2+m^2}\,t)\sim t^{-1/2}e^{-i\,\sqrt{k_\perp{}^2+m^2}\,t}\,.\quad t\to\infty
\enq

Compared to the semiclassical solution \eqref{campo WKB 0}, this defines the positive frequency states at $t\to\infty$

\beq
f_{(\infty)}^+(t)=B_{(\infty)}^+\,H_{i\,k_z}^{(2)}(\sqrt{k_\perp{}^2+m^2}\,t)\,.
\label{frecuencia positiva q=0 infinito}
\enq

Positive frequency state $f_{(\infty)}^+(t)$ is related to $f_{(0)}^+(t)$ and $f_{(0)}^-(t)$  by means of the Bogolubov transformations \citep{birrell}

\beq
f_{(\infty)}^+(t)=\alpha\,f_{(0)}^+(t)+\beta\,f_{(0)}^-(t)\,.
\label{estados bogolubov q=0}
\enq

The use of the connection formula of the Bessel functions \citep{NIST}

\beq
H_\nu^{(2)}(z)=\dfrac{e^{i\,\pi\,\nu}\,J_\nu(z)-J_{-\nu}(z)}{i\,\sin\pi\,\nu}\,,
\label{conexion bessel}
\enq

 allows us to calculate the Bogolubov coefficients $\alpha$ and $\beta$

\beq
\alpha\,A_{(0)}^+=i\,B_{(\infty)}^+\,\csc(i\,\pi\,k_z)\,,\quad \beta\,A_{(0)}^+=-i\,B_{(\infty)}^+\,e^{-\pi\,k_z}\,\csc(i\,\pi\,k_z)\,.
\enq

Then

\beq
\dfrac{|\alpha|^2}{|\beta|^2}=e^{2\pi\,k_z}\,.
\enq

Due to the orthogonality relation, the coefficients satisfy 

\beq
|\alpha|^2+|\beta|^2=1
\enq

 Therefore, the number  density of created  particles  by the evolution of the cosmological model is

\beq
n(k)=\left(\dfrac{|\alpha|^2}{|\beta|^2}-1\right)^{-1}=(e^{2\pi\,k_z}-1)^{-1}\,.
\label{numero promedio de particulas KG q=0}
\enq

This result, which is a Bose-Einstein distribution, coincides with the result by Duru  \citep{Duru} obtained with the path integral method. The dependence on the direction $k_z$ indicates that the created particle density is distributed in a uniform way along $k_z$, that is, the expansion of the universe has a privileged direction to create scalar particles.

\item Case $q=1$

In this case KG's equation solution are  Whittaker functions \eqref{campo escalar q=1}; 	the asymptotic behaviour of these functions is \citep{NIST}

For $z\to 0$

\beq
M_{\kappa\,\mu}(z)\sim z^{1/2+\mu}\,,\quad W_{\kappa\,\mu}(z)\sim \dfrac{\Gamma(2\mu)}{\Gamma(1/2+\mu-\kappa)}\,z^{1/2-\mu}\,,
\label{whittaker cero}
\enq

and if $z\to\infty$

\beq
M_{\kappa\,\mu}(z)\sim \dfrac{\Gamma(2\mu+1)}{\Gamma(1/2+\mu-\kappa)}\,e^{z/2}\,z^{-\kappa}\,,\quad W_{\kappa\,\mu}(z)\sim e^{-z/2}\,z^\kappa\,.
\label{whittaker infinito}
\enq

In the limit $t\to 0$,  the solution \eqref{campo escalar q=1} is

\beq
M_{\kappa\,,\mu}(ik_z\,t^2)\sim t^{1+i\sqrt{k_\perp^2+2\xi}}\,.
\enq 

If we compare with the asymptotic solution in $t\to 0$ by the semiclassical method we define the positive frequency states in the initial singularity as

\beq
u_{(0)}^+(t)=A_{(0)}^+\,t^{-1}\,M_{\kappa\,\mu}(i\,k_z\,t^2)\,,
\label{frecuencia positiva 0}
\enq

while negative frequency states are defined as the complex conjugate of \eqref{frecuencia positiva 0} 

\beq
u_{(0)}^-(t)=[u_{(0)}^+(t)]^*=A_{(0)}^-\,t^{-1}(e^{i\,\pi})^{1/2-\mu}\,M_{\kappa\,,-\mu}(i\,k_z\,t^2)\,.
\label{frecuencia negativa 0}
\enq

The asymptotic behaviour of the scalar field of the exact solution of the KG equation in the infinite future $t\to\infty$ for the function $W_{\kappa\,\mu}(i\,k_z\,t^2)$ corresponds to the semiclassical solution with the minus (
$-$) sign, i.e, we define the states of negative frequency in $t\to\infty$ as

\beq
u_{(\infty)}^+(t)=B_{(\infty)}^+\,t^{-1}\,W_{\kappa\,\mu}(i\,k_z\,t^2)\,.
\label{frecuencia positiva infinito}
\enq

We use Bogolubov's transformations to relate $u_{(\infty)}^+(t)$ with $u_{(0)}^+(t)\,, u_{(0)}^-(t)$
 
 \beq
u_{(\infty)}^+(t)=\alpha\,u_{(0)}^+(t)+\beta\,u_{(0)}^-(t)\,,
\label{bogolubov}
\enq

To calculate the Bogolubov coefficients, we use the Whittaker function connection formula \citep{NIST}

\beq
W_{\kappa\,\mu}(i\,k_z\,t^2)=\dfrac{\Gamma(-2\mu)}{\Gamma(1/2-\mu-\kappa)}\,M_{\kappa\,\mu}(i\,k_z\,t^2)+\dfrac{\Gamma(2\mu)}{\Gamma(1/2+\mu-\kappa)}\,M_{\kappa\,,-\mu}(i\,k_z\,t^2)\,.
\label{conexion whittaker}
\enq

Then

\beq
\dfrac{|\beta|^2}{|\alpha|^2}=e^{\pi\sqrt{k_\perp^2+2\xi}}\,.
\enq

The number  density of the  created particles is

\beq
n(k)=\left(\dfrac{|\alpha|^2}{|\beta|^2}-1\right)^{-1} =(e^{\pi\,\sqrt{k_\perp^2+2\,\xi}}-1)^{-1}\,.
\label{bose}
\enq

This expression shows that the density of particles created is the Bose-Einstein distribution, and coincides, once again, with the Duru \citep{Duru} result obtained by the path integral method.

\end{enumerate}

\section{Dirac equation }

We proceed to solve the Dirac equation in the Bianchi I model. The Dirac equation in curved space-time can be written as

\beq
\left[\tilde{\gamma}^\mu(x)\,(\partial_\mu-\Gamma_\mu)+m\right]\Psi(x)=0\,,
\enq

where $\Gamma_\mu$ the spin connections that are calculated through the expression \citep{collas}

\beq
\Gamma_\mu =\dfrac{1}{4}\,\omega_{\mu\,\alpha\,\beta}\,\gamma^\alpha\,\gamma^\beta\,,
\enq

where $\omega_{\mu\alpha\beta}$ are Ricci's rotation coefficients and $e^\mu{}_\alpha(x)$ is a tetrad that satisfies the relation

\beq
g_{\mu\nu}(x)=e_\mu{}^\alpha\,e_\nu{}^\beta\,\eta_{\alpha\beta}\,.
\enq

Dirac matrices in curved space-time are $\tilde{\gamma}^\mu(x)$ that are related to gamma matrices $\gamma^\mu$ of flat spacetime by

\beq
\tilde{\gamma}^\mu(x)=e^\mu{}_\alpha(x)\,\gamma^\alpha\,,
\enq

and comply with the anticommutation rule

\beq
\{\tilde{\gamma}^\mu\,,\tilde{\gamma}^\nu \}=2\,g^{\mu\nu}\,.
\enq

The tetrad for the metric is chosen diagonally

\beq
e^0=\dif t\,,\quad e^1=t^q\,\dif x\,,\quad  e^2=t^q\,\dif y\,\quad  e^3=t^{1-2q}\,\dif z\,,
\enq

while the spin connections  are given by

\beq
\Gamma_0=0\,,\quad\Gamma_1=\dfrac{1}{2}\,q\,t^{q-1}\,\gamma^1\,\gamma^0\,,\quad \Gamma_2=\dfrac{1}{2}\,q\,t^{q-1}\,\gamma^2\,\gamma^0\,,\quad \Gamma_3=\dfrac{1}{2}\,(1-2q)\,t^{-2q}\,\gamma^3\,\gamma^0\,,
\enq

and Dirac matrices in curved space-time are

\beq
 \tilde{\gamma}^0(x)=\gamma^0\,,\quad  \tilde{\gamma}^1(x)=t^{-q}(x)\gamma^1\,,\quad  \tilde{\gamma}^2(x)=t^{-q}(x)\gamma^2\,,\quad  \tilde{\gamma}^3(x)=t^{2q-1}(x)\gamma^3\,.
\enq

If $\Psi(x)=t^{-1/2}\,\Psi_0(x)$, the equation to be solved is

\beq
 \left[\gamma^0 \,t^q\,\partial_t  +\gamma^1\,\partial_x+\gamma^2\,\partial_y+ t^{3q-1}\,\gamma^3\,\partial_z +t^q\,m\right]\Psi_0(x)=0
 \label{dirac bianchi I}
\enq

It is a system of partial differential equations, however the fact that you have two equal and one different direction of propagation makes it difficult to separate variables.  There is a general method for separating variables in the Dirac equation in curved space-time developed by Shishkin and Villalba \citep{shishkin1}. The method consists of writing the Dirac equation in terms of the sum of two first-order differential operators

\beq
\{H\}\Psi_0(t)=\{H\}\Gamma\,\Gamma^{-1}\,\Psi_0(t)=0\Rightarrow (K_i+K_j)\Theta(x)=0\,,\quad [K_i\,,K_j]=0\,,
\label{operadores k}
\enq

where $\Theta(x)=\Gamma\Psi(x)$ is an auxiliary spinor and $\Gamma$ is a non-singular separation matrix. 

We separated variables $(t\,,z)$ from $(x\,,y)$; the operators $K_i$ are

\beq
K_1=t^q[\gamma^0\,\partial_t + t^{2q-1}\,\gamma^3\,\partial_z+m]\gamma^3\,\gamma^0\,,\quad K_2=(\gamma^1\,\partial_x +\gamma^2\,\partial_y)\gamma^3\,\gamma^0\,,
\label{operadores K}
\enq

where we choose $\Gamma=\gamma^3\,\gamma^0$. We can rewrite the Dirac equation as

\beq
(K_1+K_2)\Theta(x)=0\,.
\enq

As the Bianchi I model is homogeneous, it can be proposed a  solution of the following form

\beq
\Theta(x)=e^{i\,\mathbf{k}\cdot\mathbf{r}}\,\Theta_0(t)\,,
\label{espinor auxiliar total}
\enq

where 

\beq
\Theta_0(t)=\begin{pmatrix}
\Theta_1\\
\Theta_2
\end{pmatrix}\,,\quad \Theta_1(t)=\begin{pmatrix}
\xi_1\\
\xi_2
\end{pmatrix}\,,\quad \Theta_2(t)=\begin{pmatrix}
\xi_3\\
\xi_4
\end{pmatrix}\,.
\enq

By separating variables we have

\beq
K_1\Theta_0=-K_2\Theta_0=-k\Theta_0\,,
\enq

where $k$ is a separation constant which is obtained from the equation

\beq
K_2\Theta_0 = i(\gamma^1\,k_x+\gamma^2\,k_y)\gamma^3\,\gamma^0\,\Theta_0 =k\,\Theta_0\,.
\label{ecuacion k2}
\enq

We use the following representation of Dirac's matrices \citep{shishkin2}

\beq
\nonumber \gamma^0 =\begin{pmatrix}
0 & -\sigma^1\\
\sigma^1 & 0
\end{pmatrix}\,,\quad \gamma^1 =\begin{pmatrix}
\sigma^1 & 0\\
0 & -\sigma^1
\end{pmatrix}\,,\quad \gamma^2= -\begin{pmatrix}
\sigma^2 & 0\\
0 & \sigma^2
\end{pmatrix}\,,\quad \gamma^3=\begin{pmatrix}
0 & \sigma^1\\
\sigma^1 & 0
\end{pmatrix}\\\,.
\enq

The equation  \eqref{ecuacion k2} is equivalent to

\beq
\nonumber(k_x+i\,k_y)\Theta_1= k\,\sigma^2\,\Theta_2\\\\
\nonumber(k_x-i\,k_y)\sigma^2\,\Theta_2= k\Theta_1\,.
\enq

The solution for $\Theta_2$ in terms of $\Theta_1$ is

\beq
\Theta_2=\dfrac{k_x+\,ik_y}{k_\perp}\,\sigma^2\,\Theta_1\,,
\label{componente 2}
\enq

where $k=\sqrt{k_x{}^2+k_y{}^2}=k_\perp$ and $\sigma^2$ is the second Pauli's matrix.  Therefore the spinor $\Theta_0$ has the following structure

\beq
\Theta_0 = \begin{pmatrix}
\Theta_1\\[0.2cm]
\dfrac{k_x+i\,k_y}{k_\perp}\,\sigma^2\,\Theta_1
\end{pmatrix}\,.
\label{espinor auxiliar}
\enq

To determine $\Theta_0$ you must solve the system of equations

\beq
 \nonumber\left(\dfrac{\dif}{\dif t}+i\,k_z\,t^{2q-1}\right)\Theta_1 =\sigma^1\,(m-i\,k_\perp\,t^{-q})\,\Theta_2\\\\
 \nonumber\left(\dfrac{\dif}{\dif t}-i\,k_z\,t^{2q-1}\right)\Theta_2 =-\sigma^1\,(m+i\,k_\perp\,t^{-q})\,\Theta_1\,.
\enq

By decoupling the system we have an equation for the components of $\Theta_1(t)$ 

\beq
\nonumber \dfrac{\dif^2 \xi_{1\,,2}}{\dif t^2}&&\mp\dfrac{i\,q\,k_\perp}{m\mp i\,k_\perp\,t^{-q}}\,t^{-(q+1)}\,\dfrac{\dif \xi_{1\,,2}}{\dif t}+\\[0.3cm]
&&\nonumber +\left[m^2+k_\perp\,t^{-2q}\mp i\,k_z\,(1-2q)\,t^{2q-2}+k_z^2\,t^{4q-2}+\dfrac{k_\perp\,k_z\,q}{m\mp i\,k_\perp\,t^{-q}}t^{q-2}\right]\,\xi_{1\,,2}(t)=0\,.\\\label{ecuación componentes theta q}
 \enq
 
We proceed to find exact solutions from the above equation for  $q=0$ and $q=1$.

\subsection{Solution for $q=0$}

For $q=0$ the equation to solve is

\beq
\dfrac{\dif^2 \xi_{1\,,2}}{\dif t^2}+\left[m^2+k_\perp{}^2+k_z\,(k_z\mp i)\,t^{-2}\right]\xi_{1\,,2}(t)=0\,.
\label{Dirac q=0 con masa}
\enq

The solution are Bessel functions

\beq
\xi_{1\,,2}(t)=\sqrt{t}\,\left[A_{1\,,2}\,H_{\nu_\mp}{}^{(1)}(\sqrt{m^2+k_\perp{}^2}\,t)+B_{1\,,2}\,H_{\nu_\mp}{}^{(2)}(\sqrt{m^2+k_\perp{}^2}\,t) \right]\,,
\enq

where $A_{1\,,2}\,,B_{1\,,2}$ are integration constants and 

\beq
\nu_+=\dfrac{1}{2}+i\,k_z\,,\quad \nu_-=\dfrac{1}{2}-i\,k_z\,.
\enq

Therefore, the spinor $\Theta_1(t)$ is written as

 \beq
 \Theta_1(t)=\sqrt{t}\,\begin{pmatrix}
 A_1\,H_{\nu_-}{}^{(1)}(\sqrt{m^2+k_\perp{}^2}\,t)+B_1\,H_{\nu_-}{}^{(2)}(\sqrt{m^2+k_\perp{}^2}\,t)\\[0.3cm]
 A_2\,H_{\nu_+}{}^{(1)}(\sqrt{m^2+k_\perp{}^2}\,t)+B_2\,H_{\nu_+}{}^{(2)}(\sqrt{m^2+k_\perp{}^2}\,t) 
 \end{pmatrix}\,.
 \label{solucion q=0}
 \enq

The Dirac equation in the flat Kasner-type space-time was considered by Shishkin and Andrushkevic \citep{shishkin2} and Pimentel \citep{pim2} with $m=0$.
 
\subsection{Solution for $q=1$}

For $q=1$, the Dirac equation has exact solutions for $m=0$; the Dirac equation for this case is

\beq
\dfrac{\dif^2 \xi_{1\,,2}}{\dif t^2}+\dfrac{1}{t}\dfrac{\dif\xi_{1\,,2}}{\dif t}+\left[k_\perp{}^2\,t^{-2}\pm 2i\,k_z+k_z^2\,t^2\right]\xi_{1\,,2}(t)=0\,.
\label{dirac q=1}
\enq

Introducing the new function $\xi_{1\,,2}(t)=t^{-1/2}\,\Xi_{1\,,2}(t)$, the equation is rewritten as

\beq
\dfrac{\dif^2 \Xi_{1\,,2}}{\dif t^2}+\left[\dfrac{1}{t^2}\left(\dfrac{1}{4}+k_\perp{}^2\right)\pm 2i\,k_z+k_z^2\,t^2\right]\Xi_{1\,,2}(t)=0\,.
\enq

The solution are Whittaker functions

\beq
\xi_{1\,,2}(t)=t^{-1}\,\left[A_\pm\,M_{\pm 1/2\,,ik_\perp/2}(i\,k_z\,t^2)+B_\pm\,W_{\pm 1/2\,,ik_\perp/2}(i\,k_z\,t^2) \right]\,.
\enq

Therefore, the spinor $\Theta_1(t)$ is written as

\beq
 \Theta_1(t)&=& t^{-1}\,
\begin{pmatrix}
A_+\,M_{ 1/2\,,ik_\perp/2}(i\,k_z\,t^2)+B_+\,W_{ 1/2\,,ik_\perp/2}(i\,k_z\,t^2)\\[0.3cm]
A_-\,M_{- 1/2\,,ik_\perp/2}(i\,k_z\,t^2)+B_-\,W_{- 1/2\,,ik_\perp/2}(i\,k_z\,t^2)
\end{pmatrix}\,.
\label{solucion q=1}
\enq

\section{Production of  Spin 1/2 Particles}
 The asymptotic behaviour of the solution will define the positive and negative frequency states in $t\to 0$ and $t\to \infty$.

\begin{enumerate}

\item Case $q=0$

For $t\to 0$, positive frequency states expressed by a spinor $ \Theta_{1\;(0)}^+(t)$ are

\beq
\Theta_{1\;(0)}^+(t)=A_{(0)}^+\,\sqrt{t}\begin{pmatrix}
J_{\nu_-}(\sqrt{m^2+k_\perp{}^2}\,t)\\[0.3cm]
J_{\nu_+}(\sqrt{m^2+k_\perp{}^2}\,t)
\end{pmatrix}\,,
\enq

while negative frequency states are the complex conjugate of $\Theta_{1\;(0)}^+(t)$

\beq
\Theta_{1\;(0)}^-(t)=A_{(0)}^-\,\sqrt{t}\,\begin{pmatrix}
J_{-\nu_-}(\sqrt{m^2+k_\perp{}^2}\,t)\\[0.3cm]
J_{-\nu_+}(\sqrt{m^2+k_\perp{}^2}\,t)
\end{pmatrix}\,.
\enq

On the other hand, positive frequency states in $t\to \infty$ are given by

\beq
\Theta_{1\,(\infty)}^+(t)=B_{(\infty)}^+\,\sqrt{t}\begin{pmatrix}
H_{\nu_-}{}^{(2)}(\sqrt{m^2+k_\perp{}^2}\,t)\\[0.3cm]
H_{\nu_+}{}^{(2)}(\sqrt{m^2+k_\perp{}^2}\,t)
\end{pmatrix}\,.
\enq

The choice of the positive and negative frequency states in was based on the asymptotic behaviour of the exact solutions \eqref{solucion q=0} and \eqref{solucion q=1}, compared to the semiclassical behaviour \eqref{campo WKB 0}, \eqref{campo WKB}.

Using the connection formula of the Bessel functions, it is possible to express $\Theta_{1\,(\infty)}^+(t)$ in terms of $\Theta_{1\;(0)}^\pm(t)$ with the help of the Bogolubov transformations

\beq
\Theta_{1\;(\infty)}^+(t)=\alpha\,\Theta_{1\;(0)}^+(t)+\beta\,\Theta_{1\;(0)}^-(t)\,.
\enq

The Bogolubov coefficients

\beq
\alpha=-\dfrac{i\,C_{(\infty)}^+}{D_{(0)}^+}\,\csc(\pi\,\nu_-)\,e^{i\,\pi\,\nu_-}\,,\quad \beta =\dfrac{i\,C_{(\infty)}^+}{D_{(0)}^-}\,\csc(\pi\,\nu_-)\,.
\enq

Therefore

\beq
\dfrac{|\alpha|^2}{|\beta|^2}=e^{2\pi\,k_z}\,.
\enq

For spinors, Bogolubov's coefficients satisfy

\beq
|\alpha|^2-|\beta|^2=1\,.
\enq

The number density of created particles  is

\beq
n(k)=|\beta|^2=(e^{2\pi\,k_z}+1)^{-1}\,.
\label{creacion neutrinos q=0}
\enq

The number density of the created particles results in  a Fermi-Dirac distribution with chemical potential $\mu_{\nu_e}=0$, which means that the expansion of the universe  creates such particles thermally. 

\item Case $q=1$ with $m=0$

The semiclassical behaviour of the $\Phi$ field without  the mass term ()$m=0$) is

\beq
\Psi_k\sim\left\{\begin{array}{rcl}
e^{i\,\mathbf{k}\cdot\mathbf{r}}\,t^{\pm i\,k_\perp}\,,\quad t\to 0\label{campo dirac WKB}\\[0.3cm]
e^{i\,\mathbf{k}\cdot\mathbf{r}}\,e^{\pm i\,k_z\,t^2/2}\,,\quad t\to\infty
\end{array}
\right.
\enq

For $t\to0$, positive and negative frequency states are, respectively

\beq
\nonumber \Theta_{1\,(0)}^+(t)=C_{(0)}^+\,(e^{i\,\pi})^{1/2-\mu}\,t^{-1}\,\begin{pmatrix}
M_{ -1/2\,,-ik_\perp/2}(i\,k_z\,t^2)\\[0.3cm]
M_{+1/2\,,-ik_\perp/2}(i\,k_z\,t^2)
\end{pmatrix}\,,\quad \Theta_{1\,(0)}^-(t)=C_{(0)}^-\,t^{-1}\,\begin{pmatrix}
M_{ 1/2\,,ik_\perp/2}(i\,k_z\,t^2)\\[0.3cm]
M_{- 1/2\,,ik_\perp/2}(i\,k_z\,t^2)
\end{pmatrix}\,,\\
\enq

while for $t\to\infty$ positive frequency states are

\beq
\Theta_{1\,(\infty)}^+(t)=D_{(\infty)}^+\,t^{-1}\,\begin{pmatrix}
W_{ 1/2\,,ik_\perp/2}(i\,k_z\,t^2)\\[0.3cm]
W_{ -1/2\,,ik_\perp/2}(i\,k_z\,t^2)\,.
\end{pmatrix}
\label{estado dirac 0 +}
\enq

The spinors  $\Theta_{1\,(\infty)}^+(t)$, $\Theta_{1\,(0)}^-(t)$ are related through Bogolubov's transformations

\beq
\Theta_{1\,(\infty)}^+(t)=\alpha\,\Theta_{1\,(0)}^+(t)+\beta\,\Theta_{1\,(0)}^-(t)\,.
\label{estados infinito en 0}
\enq

Using the connection ratio of the Whittaker functions the following coefficients are obtained

\beq
\alpha=\dfrac{D_{(\infty)}^+}{C_{(0)}^+}\,\dfrac{\Gamma(-i\,k_z)}{\Gamma(-i\,k_z/2)}\,e^{\pi\,k_\perp/2}\,e^{-i\,\pi/2}	\,,\quad \beta=\dfrac{D_{(\infty)}}{C_{(0)}^-}\,\dfrac{\Gamma(i\,k_z)}{\Gamma(i\,k_z/2)}\,.
\enq

Therefore

\beq
\dfrac{|\alpha |^2}{|\beta|^2}=e^{\pi\,k_\perp}\,.
\enq

The number density of the created particles  obeys the  Fermi-Dirac distribution.

\beq
n(k)=|\beta|^2=\left(\dfrac{|\alpha|^2}{|\beta|^2} +1\right)^{-1}=(e^{\pi\,k_\perp}+1)^{-1}\,.
\label{numero particulas dirac q=1}
\enq

\end{enumerate}

\section{Final Remarks}

In this article we discuss creation of scalar and spin 1/2 particles in some LRS Bianchi I models using the semi-classical method to identify and define the positive and negative frequency states of the fields at the asymptotic limits of the $t\to0$, $t\to \infty$ expansion. For scalar particles, the same results were obtained as those  by Duru with the path integral method, while for massless spin $1/2$  particles  the result obtained is the  Fermi-Dirac distribution. This calculation corroborates that the semi-classical  method coincides with the more sophisticated of path integrals to define the vacuum state in general relativity, which makes it  efficient and accessible in many cases of interest.

\section{\bf{Acknowlegements}} This work was partially supported by PRODEP UAM-I-CA-43 and F. P. by CONACYT grant 706699.

\end{document}